\documentclass[%
 aip,
 amsmath,amssymb,
 reprint,%
]{revtex4-1}

\usepackage{graphicx}
\usepackage{dcolumn}
\usepackage{bm}
\usepackage{xcolor}

\usepackage[utf8]{inputenc}
\usepackage[T1]{fontenc}
\usepackage{mathptmx}
\usepackage{etoolbox}

\makeatletter
\def\@email#1#2{%
 \endgroup
 \patchcmd{\titleblock@produce}
  {\frontmatter@RRAPformat}
  {\frontmatter@RRAPformat{\produce@RRAP{*#1\href{mailto:#2}{#2}}}\frontmatter@RRAPformat}
  {}{}
}%
\makeatother
\begin{document}

\preprint{AIP/123-QED}

\title[Relative permeability as a stationary process]{Relative permeability as a stationary process: energy fluctuations in immiscible displacement}
\author{James E McClure}
  \email{mcclurej@vt.edu}
\affiliation{National Security Institute, Virginia Polytechnic Institute \& State University, 
Blacksburg, Virginia, 24060, USA}

\author{Ming Fan}%
\affiliation{Computational Sciences and Engineering Division, Oak Ridge National Laboratory, Oak Ridge, Tennessee, 37830, USA}%

\author{Steffen Berg}
\affiliation{%
Shell Global Solutions International B.V.
Grasweg 31,
1031HW Amsterdam,
The Netherlands
}%

\author{Ryan T. Armstrong}
\affiliation{%
School of Minerals \& Energy Resources Engineering, University of New South Wales, Kensington, NSW 2052, Australia
}

\author{Carl Fredrik Berg}
\affiliation{%
PoreLab, Department of Geoscience and Petroleum, Norwegian University of Science and Technology (NTNU), Trondheim, Norway
}

\author{Zhe Li}%
\affiliation{ 
Research School of Physics, The Australian National University, Canberra, ACT 2601, Australia
}%

\author{Thomas Ramstad}%
\affiliation{ 
Equinor ASA, Arkitekt Ebbells veg 10, Rotvoll, NO-7005, Trondheim, Norway
}%

\date{\today}

\begin{abstract}
Relative permeability is commonly used to model immiscible fluid flow through porous materials. In this work we derive the relative permeability relationship from conservation of energy, assuming that the system to be non-ergodic at
large length scales and relying on averaging in both space and time to homogenize the behavior. Explicit criteria are obtained to define stationary conditions:
(1) there can be no net change for extensive measures of the system state over the time averaging 
interval; (2) the net energy inputs into the system are zero, meaning that the net rate of work done on the system must balance with the heat removed; and (3) there is no net work performed due to the contribution of internal energy fluctuations. Results are then evaluated based on direct numerical simulation. Dynamic connectivity is observed during steady-state flow, which is quantitatively assessed based the Euler characteristic. We show that even during steady-state flow at low capillary number ($\mathsf{Ca}\sim1\times10^5$), typical flow processes will explore multiple connectivity states. The residence time for each connectivity state is captured based on the time-and-space average. The distribution for energy fluctuations is shown to be multi-modal
and non-Gaussian when terms are considered independently. However, we demonstrate that their
sum is zero. Given an appropriate choice of the thermodynamic driving force, we show that the conventional relative permeability relationship is sufficient to model the energy dissipation in systems with complex pore-scale dynamics that routinely alter the structure of fluid connected pathways.
\end{abstract}

\maketitle

Relative permeability provides the fundamental basis to model immiscible fluid flow through permeable media. While the concept of relative permeability has been well-established for decades, an adequate theoretical justification has remained elusive.
Canonical arguments are based on the form of Darcy's law and established on a phenomenological basis.  The lack of theoretical support means that experimental measurements must be performed and interpreted without a complete picture of the physical assumptions needed to observe the intended behavior \cite{Berg_Unsal_etal_2021}.
Efforts to derive the conventional relative permeability from first-principles have focused primarily on momentum conservation \cite{muskat1936,wyckoff1936,leverett1941,buckley1942}. 
While the single fluid version of Darcy's law can be derived on this basis \cite{whitaker1986flowa}, 
it is well-known that volume-averaging for momentum transport in a multiphase system leads to cross-coupling forms that do not align with the conventional concept of the relative permeability \cite{whitaker1986flowb,rose1988,flekkoy1999}. 
In this work, we argue that the conventional relative permeability is best understood from the perspective of energy conservation. Considering the non-equilibrium thermodynamics from this perspective, we show that the validity of the steady-state relative permeability can be established based on a stationary fluctuation condition. 

The conceptual basis for the relative permeability considers the situation where each fluid forms a connected pathway through the material, which are the conduits through which fluid flow occurs \cite{muskat1936flow, blunt2017multiphase}. As the fluid
volume fraction changes connected pathways break down, and fluids become immobilized within the pore space \cite{corey1954, brooks1964, land1968,stone1970}. Mobilization of these trapped fluids only occurs as the capillary number increases, which is defined as the ratio of viscous
to capillary forces. At low capillary number, the interface configuration is often assumed to be static due to the dominance of capillary forces. As the flow rate increases, the arrangement of fluids is sensitive to the capillary number and viscosity ratio \cite{fulcher1985effect,avraam1995flow,li2005pore,ramstad2012relative, fan2018,fan2019}. The wetting characteristics of the porous material are also known to influence the relative permeability \cite{anderson1987,blunt2001flow,oren2003,valvatne2004,spiteri2008,zhao2016,fan2020,guo2020}. These effects can be demonstrated to alter the pore-scale arrangement of 
fluids, including the topological structure of the fluids \cite{herring2013effect,armstrong2016pre,schluter2016pore, mcclure2018geometric,armstrong2019porous,sun2020}. 
Recent work has also established that changes to the fluid topology occurs during flow at lower flow rates, and that the flow pathways can be intermittent for some fluid saturation values \cite{reynolds2017pnas,spurin2019intermittent}.

Laboratory measurement of the relative permeability is most typically performed based on fractional flow experiments \cite{geffen1951experimental,osoba1951laboratory,richardson1952laboratory,honarpour1988,bennion2008drainage,krevor2012}. 
In a typical steady-state experiment, two fluids are co-injected at the inlet and flow through the sample until a desired steady state fractional flow condition is met, such as the pressure and saturation become time-independent. Through monitoring the pressure gradients and flow rates, the relative permeability can be determined based on the extended version of Darcy’s law \cite{bear2012}.  The relative permeability curves can be obtained with a step-wise manner by changing the fluid flow rates and repeating the measurements at saturation equilibrium states. In many situations, the core flooding experiments may only generate quasi steady-state flows,
such that the fluid saturation and pressure values fluctuate at timescales that are large compared with the timescale at which experimental measurements are performed \cite{persoff1995,reynolds2015characterizing,sorop2015relative,gao2017x,lin2018imaging,gao2019pore,lin2019minimal,wang2019obtaining,clennell2020improved,gao2020pore,rucker2021}.

Slow fluctuations indicate that the timescale for experimental observations is too fast compared with the relaxation dynamics of the system \cite{schluter2017relaxation}, in which case the ergodic hypothesis may fail \cite{doi:10.1080/00018738200101438}. Ergodicity is formally achieved in the limit of infinite time, which is sufficient to describe equilibrium as well as many near-equilibrium systems \cite{doi:10.1073/pnas.17.2.656,doi:10.1073/pnas.18.1.70}. Since ergodicity is fundamentally linked with equipartition of energy, the timescale to achieve
ergodic conditions is fundamentally linked with timescale for thermal mixing. 
In an ergodic system, fluctuation statistics will be Gaussian, and spatial, temporal and ensemble averages will be equivalent. Previously, it has been suggested that
ergodicity breaking should be expected when cooperative mechanisms  move mass and energy more rapidly than the timescale for thermal mixing
\cite{mcclure2020capillary}. As fluids travel through the porous medium, complex changes to the fluid configuration occur because of the capillary dynamics within the system. The fluid displacement often involves a sequence of pore-scale events, which can be categorized as reversible displacement events known as isons and abrupt displacements associated with disruption to the interfaces referred to as rheons \cite{morrow1970}. 
In the pore-filling events, such as Haines jump, the local capillary pressure of the fluid meniscus suddenly drops due to the decrease in the meniscus curvature, and then fluids invade the neighboring pores in an accelerated manner until the local capillary pressure is balanced again by the fluid pressure difference. As a result of these fluid burst events, a cascade \cite{armstrong2014subsecond} of energy dissipation between the invading and defending fluids emerge, which contributes to continuous variations in fluid pressure and saturation \cite{moebius2012interfacial,berg2013,rucker2015connected}. 
The timescale for these events are directly linked with the observed fluctuations.

In this study, we derive the standard relationship for relative permeability based on time-and-space averaging and conservation of energy.
Explicit consideration for both time and space has been previously developed,
from without full consideration of global energy conservation \cite{Kalaydjian_1987}. We show that the standard form of Darcy's law can be obtained by considering stationary energy balance with minimal assumptions. We define stationary conditions to mean the extensive properties of the system must be constant, and integrate the general expression for conservation of energy to obtain an explicitly scale-dependent representation. The scale-dependent representation can be directly linked to physical experiments, such that valid transport coefficients may be obtained to characterize the dissipation without depending on the existence of a representative elementary volume (REV). We show that the timescale to obtain stationary conditions 
is associated with an energy fluctuation constraint, providing a clear path to 
obtain generalizations of Darcy's law that can be applied in fluctuating systems. 
Simulation results validate the approach for two-fluid flow in porous media,
identifying key terms and their relative importance in practical situations
as a measure of experimental uncertainty. 

\section{Energy Dynamics of a Stationary System}

In this work, we derive the standard equation for relative permeability on the basis of conservation of energy. We consider the multiphase extension of Darcy's law 
in the form,
\begin{equation}
 \bar{\mathbf{q}}_i = - \frac{k^r_i \mathbf{\mathsf{K}}}{\mu_i}  \cdot \nabla \bar{p} 
 \label{eq:Darcy}
\end{equation}
where $\mathbf{\mathsf{K}}$ is the permeability tensor, $\bar{p}$ is the average pressure, $\mu_i$ is the dynamic viscosity for fluid $i$, $\bar{\mathbf{q}}_i$ is the associated volumetric flow rate and $k^r_i$ is the relative permeability. In the results that follow we will derive 
Eq. \ref{eq:Darcy} from first principles, providing scale 
consistent definitions for all quantities in the process. In particular
we will show that the average pressure $\bar{p}$ must be the volume-averaged total pressure, including contributions due to both fluids. 
Eq. \ref{eq:Darcy} can be considered
as a linear phenomenological equation to predict the flow rate based on the thermodynamic driving force $-\nabla \bar{p}$, with the coefficient tensor
${k^{r}_i \mathbf{\mathsf{K}}}/{\mu_i}$ relating to the rate of energy dissipation in the system. 
Efforts to derive the conventional relative permeability form based on volume-averaging
theory have been largely unsuccessful, instead leading to cross-coupling forms
\cite{whitaker1986flowb,rose1988,flekkoy1999}. These forms are intuitive from the perspective of tangential stress boundary continuity in the picture of connected pathways with adjacently flowing liquid phases. However, significant experimental challenges are introduced based on cross-coupling, since more information is needed to disambiguate the additional two 
coefficients needed to form the approximation. Since multiphase flow experiments are expensive to perform, the standard two-phase Darcy form is attractive based on the fact that all necessary coefficients can be inferred from a single experiment. In this paper we demonstrate that cross-coupling forms can be avoided based on on an appropriate choice of the thermodynamic driving force. We directly derive Eq. \ref{eq:Darcy} based on the conservation of energy in a stationary system. A set of three constraints are derived to establish meaning of stationary conditions within this context. Ultimately, these reflect the fact that the rate of work done on the system must balance with the dissipated heat for a stationary process. Critically, we do not assume that the system is ergodic at the Darcy scale in our derivation \cite{mcclure2020capillary,mcclure2021thermodynamics}. 

Consistent with the non-equilibrium theory developed by Onsager, the result will be determined from a flux-force product that is obtained under stationary conditions \cite{PhysRev.37.405}. The derivation is developed as a scale-dependent relationship, without assuming that a representative elementary volume (REV) exists. The existence of a REV has been called into question due to the significant length scale heterogeneity that is typical for geologic materials as well as due to the accompanying scaling behavior for fluid cluster sizes \cite{PhysRevE.88.033002}.
Based on our approach, relative permeability coefficients may be obtained for any system
based on the measured rate of dissipation. If a REV exists, this corresponds to situations where these coefficients become scale invariant. Stationary conditions are enforced  by requiring that there is no net energy input into the region when considered over a particular timescale, which can be considered as a particular quasi-ergodic requirement. We consider a system with constant spatial domain $\Omega \in \Re ^3$, which has a boundary $\Gamma$ that encloses a volume $V$. While from the mathematical perspective the subsequent theoretical development is valid for arbitrary $\Omega$, it is nevertheless useful to consider this entity as being physically significant, e.g. to represent a physical sample used in a relative permeability experiment. For the situation where $\Omega$
corresponds to a physical sample, the approach can be considered as an explicit 
framework for averaging. Since an integral over $\Omega$ is a definite integral, 
\begin{equation}
    \int_{\Omega} f dV = \sum_i \int_{\Omega_i} f dV 
    \label{eq:subset}
\end{equation}
where $\Omega_i \cap \Omega_j = 0$ for $j\neq i$ are non-intersecting subsets of $\Omega$. The union of $\Omega_i$ forms $\Omega$, i.e. the subsets represent a spatial decomposition of the system. For immiscible flow through porous media, three subsets are sufficient: $\Omega_w$  corresponds to the part of space occupied by the wetting fluid, $\Omega_n$ to the non-wetting fluid and $\Omega_s$ to the solid, such that $\Omega = \Omega_s \cup \Omega_w \cup \Omega_n$. It is important to note that the structure of any subset will not be independent from the others. Measures of the system behavior that are dependent on the structure of subsets must obey additional constraints that do apply to the global system. It is not trivial to formulate the geometric constraints, since any subset $\Omega_i$ can undergo discontinuous transformations \cite{geometric_state_euler_2020}. It is therefore advantageous to begin from a global perspective, where no discontinuous transitions are possible, and only later segregate averaged terms based on the spatial structure of the system. Furthermore, averaging in both time and space provides a mechanism to smooth discontinuities that arise due to geometric evolution. This is necessary based on the fact that recent experiments demonstrate that topological structure of fluids can change during steady-state flow \cite{spurin2019intermittent}. We now consider the development of a scale-dependent expression based on integration in both time and space, consistent with previous developments \cite{mcclure2021thermodynamics,mcclure2020capillary}.

We begin from basic conservation principles for mass, momentum and energy. Furthermore,
we rely on the following principle of scale consistency: the basic meaning for additive (i.e. extensive) quantities must not be altered based on scale transformation. Quantities such as mass, momentum, and energy have precise definitions and are always integrable. Scale transformations that alter the meaning of these fundamental quantities should be considered to be invalid. Furthermore, forces must also be additive to maintain scale consistency with respect to Newton's second law of motion. This imposes a constraint on how to define the stress tensor across scales based on integration. Provided that scale consistency is observed, the transport coefficients will properly encode the rate of dissipation; the particular values for transport coefficients may depend on the scale of observation. 

Consider a thermodynamic system of $\mathcal{N}$ components such that the internal energy of the microscopic thermodynamic system is given at equilibrium by $U(S,N_k)$, 
where $S$ is the entropy and $N_k$ is the number of molecules for component
$k=1,\ldots,\mathcal{N}$. The Gibb's equation is determined from Euler's homogeneous function 
theorem \cite{JONGSCHAAP20015},
\begin{equation}
    U = TS + \mu_k N_k \;.
    \label{eq:Gibbs}
\end{equation}
The intensive thermodynamic conjugate variables are defined as 
\begin{equation}
      T = \Big(\frac{\partial U}{\partial S}\Big)_{N_k}\;, \quad
      {\mu}_k = \Big(\frac{\partial U}{\partial N_k}\Big)_{S,N_j \neq i}\;.
\end{equation}
In a spatially heterogeneous system, the chemical potentials $\mu_k$ can vary
within the system, for example due to the surface excess potential at interfaces. 
In many situations, spatial heterogeneity can be treated based on the Gibbs adsorption isotherm,
leading to the forms derived in reference \cite{mcclure2020capillary}. 
The microscopic contribution to equilibrium internal energy is due to the rate of energy 
accumulation  based on internal stresses, e.g. due to capillary stresses
\cite{PhysRevE.102.033113}. Non-equilibrium effects also arise, and the
energy dynamics due to these effects can include both reversible and irreversible contributions based on the interplay of viscous and capillary forces. For example, a Haines jump will produce a burst in viscous energy dissipation that is accompanied by a reversible increase in surface energy. To disambiguate the reversible and irreversible contributions, the energy dynamics must be considered from a more detailed level, incorporating the thermodynamic description.

In statistical approaches, ergodicity is often assumed to hold explicitly 
at large length scales, since the system will then have attractive statistical properties \cite{PhysRevE.95.023116}. We do not make this assumption here. 
Instead, we obtain a larger scale representation by directly integrating the 
system dynamics in both time and space, assuming ergodicity will hold 
only at sufficiently small length scales. As in previous development \cite{mcclure2021thermodynamics}, we introduce the ergodic volume $\mathcal{V}$ as the region of space where local equilibrium can be assumed. At any scale larger than $\mathcal{V}$, we do not assume that an ensemble average is equivalent to a temporal or spatial average. 
Due the fact that the mean-squared distance associated with mass diffusion scales linearly with time, mixing will occur rapidly for sufficiently small $\mathcal{V}$, but can occur very slowly based on the volume of the larger system, $V > \mathcal{V}$.
The effect at larger scales is clear from considering the timescale for fast pore-scale events such as Haines jumps, because that these events move mass faster than the energy micro-states can mix based on the diffusive timescale \cite{mcclure2020capillary}.
At the small scale of $\mathcal{V}$ we assert that any deviation from equilibrium are locally linear such that quasi-ergodic conditions apply. We can therefore define quantities on a per-unit-volume basis using $\mathcal{V}$ to define the size of a local reference volume. For example, the mass density for component $k$ and total mass density are given by
\begin{equation}
    \rho_k = \frac{m_k N_k}{\mathcal{V}}\;, \quad \rho = \sum_k \rho_k
\end{equation}
where $m_k$ is the mass for a molecule of type $k$, and $N_k$ is the number of molecules
contained within a spherical reference volume of size $\mathcal{V}$. The number of
molecules $N_k$ may vary in both space and time, but $\mathcal{V}$ is independent of
both time and space. 

Away from equilibrium, the system must obey the general expression for conservation of energy,
which is derived from standard conservation principles,
\begin{equation}
\frac{\partial }{\partial t} \frac{U}{\mathcal{V}}
+ \nabla \cdot \Big(\frac{\mathbf{u} U}{\mathcal{V}} \Big)
-\bm{\sigma} : \nabla \mathbf{u} - \nabla \cdot \mathbf{q}_h - {q}_s  = 0\;,
\label{eq:energy-i}
\end{equation}
where $\mathbf{u}$ is the rate of mass flux, $\bm{\sigma}$ is the stress tensor, and 
$\mathbf{q}_h$ is the heat flux. For immiscible fluid flow, we assume that there are no significant contributions due to chemical reactions, so that body sources for heat can be ignored, ${q}_s=0$. The constant reference volume $\mathcal{V}$ corresponds to a sufficiently small region where the ergodic hypothesis is assumed to hold.  This means that deviations from equilibrium should be locally linear across $\mathcal{V}$. 

The stress tensor $\bm{\sigma}$ can be decomposed into the mean and deviatoric components, 
\begin{equation}
    {\bm{\sigma}} = -{p} \bm I + {\bm{\tau}} \;,
\end{equation}
where $p$ is the pressure and $\bm{\tau}$ is the deviatoric stress tensor. 
 Eq. \ref{eq:energy-i} can be rearranged to find the form
\begin{eqnarray}
\frac{\partial }{\partial t} \frac{U}{\mathcal{V}}
+\nabla \cdot \Big[\mathbf{u} \Big( \frac{U}{\mathcal{V}}+ p \Big)- \mathbf{q}_h\Big] 
-\mathbf{u} \cdot \nabla p
-\bm{\tau} : \nabla \mathbf{u} = 0
\label{eq:energy-ii}
\end{eqnarray}
Let $\Lambda$ be a time interval with duration $\lambda$. Conservation of energy for the region as a whole is 
given by the integral of \ref{eq:energy-ii}, 
\begin{eqnarray}
\int_{\Lambda}\int_{\Omega} \Bigg(
\frac{\partial }{\partial t} \frac{U}{\mathcal{V}}
+\nabla \cdot &&\Big[\mathbf{u} \Big( \frac{U}{\mathcal{V}} + p \Big)- \mathbf{q}_h\Big]
- \nonumber \\ && \mathbf{u} \cdot \nabla p
-\bm{\tau} : \nabla \mathbf{u} \Bigg) dV dt = 0
\label{eq:energy-iii}
\end{eqnarray}
To guide the averaging strategy used for Eq. \ref{eq:energy-iii}, we first identify 
extensive properties of the system and define averages for each quantity. 
Scale-consistent definitions are established to average the non-equilibrium dynamics over a larger region of time and space based upon the time-and-space averaging operator, 
\begin{equation}
\big<f\big > \equiv \frac{1}{\lambda \mathcal{V} }\int_{\Lambda} \int_\Omega fdVdt\;.
\end{equation}
Note that this definition is distinct from an ensemble average, since it
is fully deterministic and accounts for the complete path of the system over
the regions of space $\Omega$ and time $\Lambda$.
The extensive thermodynamic quantities from Eq. \ref{eq:Gibbs} are $U$, $S$ and $N_k$. 
Since these quantities are additive, 
\begin{equation}
\bar{U} \equiv \big< U\big>\;, \quad
\bar{S} \equiv \big< S\big>\;, \quad
\bar{N_k} \equiv \big< N_k\big>\;.
\end{equation}
The intensive thermodynamic quantities are then defined 
\begin{equation}
\bar{T} \equiv \frac{\big< T S\big>}{\bar{S}}\;, \quad
\bar{\mu_k} \equiv \frac{\big< \mu_k N_k\big>}{\bar{N_k}}\;.
\end{equation}
The total mass and total momentum are also additive, and are necessarily 
defined as
\begin{eqnarray}
\bar{M} &=& \big< m_k N_k \big>
\label{eq:mass-total}
\\
\bar{\mathbf{j}}_M &=&  \big< m_k N_k \mathbf{u} \big>
\label{eq:mass-flux-total}
\end{eqnarray}
where $m_k$ is the mass of molecule of type $k$. 
The average velocity corresponds to the rate of mass flux, which is given by the ratio
\begin{equation}
\bar{\mathbf{u}} =  \frac{\bar{\mathbf{j}}_M}{\bar{M}}  \;.
\label{eq:mass-flux-average}
\end{equation}
Since volume is an additive measure, the definition of volume fraction for any entity 
must be scale consistent. Note that while the domain $\Omega$ is independent of time, the sub-region occupied by either fluid $\Omega_i$ for $i\in \{w,n\}$ may be time dependent.
The time-averaged saturation and porosity are given by 
\begin{equation}
\bar{s}_i = \frac{1}{\bar{\phi} \lambda V} \int_{\Lambda}\int_{\Omega_i} \;dV dt \;, \quad
\bar{\phi} = \sum_{i\in\{w,n\}} \frac{1}{\lambda V} \int_{\Lambda}\int_{\Omega_i} \;dV dt \;.
\end{equation}
Finally, noting that forces are additive, the average pressure should be obtained
as an integral of the microscopic pressure field. If some sub-region of $\Omega$ is  
occupied by solid, then the pressure should be weighted based on the 
porosity $\bar{\phi}$ to correspond with the region of space occupied by fluids. 
The average pressure is therefore given by
\begin{eqnarray}
\bar{p} = \frac{1}{\bar{\phi} \lambda V} \int_{\Lambda} \int_{\Omega} p \; dV dt \;,
\label{eq:pressure-definition}
\end{eqnarray}
where $V$ is the volume of $\Omega$.

Based on these definitions, we now proceed to express Eq. \ref{eq:energy-iii} in terms
of these quantities. A simplified form will be obtained when the energy dynamics are 
stationary over the timescale associated with $\Lambda$. First we consider the energy
inputs and outputs, which are boundary fluxes of heat, work and internal energy.
These fluxes are easily identified by applying the Gauss divergence theorem, 
\begin{equation}
\int_{\Omega} 
\nabla \cdot \Big[\mathbf{u}\Big( \frac{U}{\mathcal{V}}  + p\Big) 
- \mathbf{q}_h \Big]dV = 
\int_{\Gamma} 
\Big[\mathbf{u} \Big( \frac{U}{\mathcal{V}} + p \Big) - \mathbf{q}_h \Big]\cdot \bm{dA} \;,
\label{eq:energy-input}
\end{equation}
where the vector-valued surface element $\bm{dA}$ is a normal vector that points outward from the 
boundary surface $\Gamma$. 
For a stationary process the energy inputs must be zero, since otherwise there will be a net accumulation of energy within the system. The first constraint on stationary conditions is therefore
associated with the energy inputs, 
\begin{equation}
\int_{\Lambda} \int_{\Gamma} 
\Big[\mathbf{u} \Big( \frac{U}{\mathcal{V}} + p \Big) - \mathbf{q}_h \Big]\cdot \bm{dA} dt = 0\;.
     \label{eq:stationary-constraint-i}
\end{equation}
The time interval $\Lambda$ must be sufficiently large that any fluctuations to the energy 
inputs will cancel.

The internal work due to the pressure gradient can be formally averaged to obtain 
an averaged expression in terms of $\bar{\mathbf{u}}$ and $\bar{p}$,
\begin{eqnarray}
\int_{\Lambda}\int_{\Omega} \mathbf{u} \cdot \nabla p  \;dV dt
&=& \int_{\Lambda}\int_{\Omega} \big(\bar{\mathbf{u}} + \mathbf{u}^\prime \big) \cdot \nabla p\;dV dt \;  \nonumber \\
&=& \bar{\mathbf{u}} \cdot  \nabla  \int_{\Lambda}\int_{\Omega}  p  \; dV dt 
+ \int_{\Lambda}\int_{\Omega}  \mathbf{u}^\prime \cdot \nabla p   \; dV dt \;  \nonumber \\
&=& \lambda V \; \bar{\mathbf{u}} \cdot \nabla  \big( \bar{\phi} \bar p \big)
+ \int_{\Lambda}\int_{\Omega}  \mathbf{u}^\prime \cdot \nabla p   \; dV dt.
\label{eq:energy-iv-lhs-average-i}
\end{eqnarray}
The velocity deviation is defined as $\mathbf{u}^\prime = {\mathbf{u}} - \bar{\mathbf{u}}$,
which will vary in time and space based on the pore structure and any dynamic changes
to the flow field \cite{PhysRevE.102.033113}.

The internal energy dynamics are averaged as previously described 
\cite{mcclure2020capillary,mcclure2021thermodynamics},
\begin{eqnarray}
\frac{1}{\lambda \mathcal{V}} \int_{\Lambda}\int_{\Omega} \frac{\partial U}{\partial t} dV dt
&=& \bar{T} \frac{\partial \bar{S} }{\partial t} 
+ \bar{\mu_k} \frac{\partial \bar{N}_k }{\partial t} 
\nonumber \\ &&
-\Big< S \frac{\partial T^\prime}{\partial t} \Big>
-\Big< N_k \frac{\partial \mu_k^\prime}{\partial t} \Big> \;.
\label{eq:thermodynamics}
\end{eqnarray}
Deviation terms are defined based on the difference between the local value for intensive thermodynamic
properties with their time-and-space averaged values,
\begin{equation}
     T^\prime = T - \bar{T}\;, \quad
     \mu_k^\prime = \mu_k - \bar{\mu}_k\;.
     \label{eq:intensive-deviation}
\end{equation}
Stationary conditions further require that there be no change to extensive thermodynamic state variables,
\begin{equation}
\frac{\partial \bar{S} }{\partial t} = 0\;, \quad
\frac{\partial \bar{N}_k }{\partial t} = 0\;.
     \label{eq:stationary-constraint-ii}
\end{equation}
Note that the average entropy $\bar{S}$ should be constant within the sample
based on this constraint. The implication is that any entropy produced due to irreversible events is removed based on the heat flux $\nabla \cdot \mathbf{q}_h$. 
This is captured based on the first constraint stated in Eq. \ref{eq:stationary-constraint-i}. 

Combining Eqs. \ref{eq:energy-iv-lhs-average-i} and \ref{eq:thermodynamics},
along with the constraints for stationary dynamics given in Eqs. \ref{eq:stationary-constraint-i}
and \ref{eq:stationary-constraint-ii} and rearranging terms, we can obtain
a form where the viscous energy dissipation is linked to a flux-force product,
\begin{eqnarray}
-\bar{\mathbf{u}} \cdot \nabla \big( \bar{\phi} \bar{p} \big) 
= \frac{1}{\lambda V}\int_{\Lambda}\int_{\Omega}
\bm{\tau} : \nabla \mathbf{u}\;
dV dt  
\label{eq:energy-v}
\end{eqnarray}
This form is obtained only in the situation where the work due to the fluctuation
terms cancels
\begin{eqnarray}
\frac{1}{\lambda V}\int_{\Lambda}\int_{\Omega}
\Bigg(  
\mathbf{u}^\prime\cdot \nabla p
+ \frac{S}{\mathcal{V}} \frac{\partial T^\prime}{\partial t}
+ \frac{N_k}{\mathcal{V}}\frac{\partial \mu_k^\prime}{\partial t} 
\Bigg) dV dt = 0 \;. 
\label{eq:fluctuation}
\end{eqnarray}
This equation implies that there must be no net work done due to fluctuations in the pressure and chemical potential when considered over the time interval $\Lambda$. Eq. \ref{eq:fluctuation}
can be considered as an alternative expression for Onsager's assumption of molecular reversibility
effectively stating a criterion for detailed balance in the system based purely on classical thermodynamic concepts. 
The advantage of this form is that it is straightforward to study from the perspective of either mesoscopic or continuum scale simulation techniques. Fluctuations associated with heterogeneity related to intensive thermodynamic properties are in equilibrium when no net work is performed due to their effects. Note that this is a less scrict criterion as compared to detailed balance, which would imply statistical indistinguishability. Based on these results, conditions for the validity of Onsager theory
can be established based for systems that may be non-ergodic when considered at the timescale
where measurements are taken. An alternative interpretation is that the finite time interval 
$\Lambda$ can also be considered to define the length of time needed for the system to behave
as an ergodic system, which can be evaluated quantitatively based on
Eqs. \ref{eq:stationary-constraint-i}, \ref{eq:stationary-constraint-ii} and \ref{eq:fluctuation}. In the Results section we will evaluate Eq. \ref{eq:fluctuation} directly based on simulations of immiscible displacement through porous media. 

\subsection{Darcy's Law (Single phase)}

The previous result can be immediately applied to derive Darcy's law. The right-hand side of Eq. \ref{eq:energy-v} accounts for the heat generated due to viscous dissipation. Considering a Newtonian fluid with constant viscosity $\mu$, 
\begin{equation}
 \int_{\Lambda}\int_{\Omega}
\bm{\tau} : \nabla \mathbf{u}\; dV dt  = 
 \mu \int_{\Lambda} \int_{\Omega}
\mathbf{E}  : \nabla \mathbf{u}\; dV dt  
\label{eq:Darcy-viscosity}
\end{equation}
where the strain rate tensor is
$\mathbf{E} = \frac{1}{2}\big[ \nabla\mathbf{u} 
+ \big(\nabla\mathbf{u} \big)^T \big]$. Note that $\mu$
can only be taken outside of the integral if it is constant
over the region of space and time specified by $\Omega$ and $\Lambda$. 
This assumption will fail in a system with compositional heterogeneity,
and also in situations where the stress depends non-linearly on the
strain rate tensor. 
To derive Darcy's law, we use the fact that the 
positive heat is generated due to viscous dissipation, leading to an inequality
\begin{eqnarray}
-\bar{\mathbf{u}} \cdot \frac{\nabla \big( \bar{\phi} \bar{p} \big)}{\mu}
\ge 0
\label{eq:darcy-i}
\end{eqnarray}
As the  driving force $\nabla ( \bar{\phi} \bar{p}) $ goes to zero
the corresponding flux $\bar{\mathbf{u}}$ will go to zero also;
any stationary fluctuations must obey Eq. \ref{eq:fluctuation}
due to conservation of energy in the system. This tells us the timescale
required to homogenize the system dynamics. For sufficiently small driving
forces, causality leads us to approximate the flux based on a linear expansion in terms of the driving force
\begin{eqnarray}
\bar{\mathbf{u}} = -\frac{\mathbf{\mathsf{L}}}{\mu} \cdot
\nabla \big( \bar{\phi} \bar{p} \big)
\label{eq:darcy-ii}
\end{eqnarray}
where the tensor $\mathbf{\mathsf{L}}$ contains linear 
phenomenological coefficients, consistent both with the approach of Onsager
as well as results obtained using volume-averaging techniques \cite{Whitaker_Darcy_1986}. The phenomenological coefficient tensor is related 
to the absolute permeability tensor for the material,
\cite{doi:10.1021/es049728w,GRAY20061745}
\begin{equation}
\mathbf{\mathsf{K}} = \bar{\phi}^2 \mathbf{\mathsf{L}} \;.
\end{equation}
For a system with homogeneous porosity, Darcy's law is obtained in the standard
form
\begin{equation}
\bar{\mathbf{q}} = -\frac{\mathbf{\mathsf{K}}}{\mu} \cdot
\nabla  \bar{p}  \;,
\label{eq:darcy-standard}
\end{equation}
where the flux $\bar{\mathbf{q}} = \bar{\phi}\bar{\mathbf{u}}$.
For a Stokes flow, the velocity field depends only on the geometry, and $\mathbf{\mathsf{K}}$ is directly determined
from this based on the right-hand side of Eq. \ref{eq:Darcy-viscosity}.
Since the Stokes solution does not depend on time, there is no advantage to
time-and-space averaging to derive the single phase Darcy's law. 
We expect that a single fluid flow will satisfy macroscopic ergodic
requirements. However, this breaks down in immiscible displacement,
due to slow fluctuations that are caused by pore-scale events \cite{rucker2021,mcclure2020capillary}. 
While the form of Eqs. \ref{eq:energy-v} and
\ref{eq:fluctuation} remains valid for two-fluid systems, the viscosity cannot be treated as in  Eq. \ref{eq:Darcy-viscosity}, and the permeability 
will consequently depend on the spatial arrangement of fluids. 
Even in cases where the viscosity varies in space and time, this does not present a significant difficulty to derive Eq. \ref{eq:darcy-ii}; an
inequality as given in Eq. \ref{eq:darcy-i} can still be obtained 
since $\mu>0$. However, the effective permeability will depend on 
both the spatial and temporal structure of the fluid flow field, and 
the thermodynamic driving forces must be considered carefully. 

\subsection{Multiphase extension of Darcy's law}

The standard conceptual model for the relative permeability considers the situation where two
fluids flow simultaneously through a porous medium through separate flow pathways.
Emerging models also consider the situation where fluid pathways rearrange dynamically.
Our model accounts for both possibilities, since any dynamics will be captured explicitly
based on the time average. To account for the flow rate of each fluid component, 
it is necessary to represent the mass flux separately for each fluid. 
To obtain Eq. \ref{eq:Darcy}, flow behavior must be associated with each fluid, which is
accomplished by applying a subset operation to sub-divide the integrals given in Eqs.
\ref{eq:mass-total} -- \ref{eq:mass-flux-average}. The sub-setting operation is applied to affiliate the mass and momentum with a particular phase based on the location of the Gibb's dividing surface within the microscopic system \cite{adamson97}.
Since the flow rate is determined based on an integral over the spatial region
$\Omega$, Eq. \ref{eq:mass-flux-total} be decomposed into components associated with the part of space 
occupied by water $\Omega_w$ and the  part of space occupied by the non-wetting fluid $\Omega_n$
\begin{eqnarray}
\bar{M} \bar{\mathbf{u}} &=&  \bar{M}_w \bar{\mathbf{u}}_w + \bar{M}_n \bar{\mathbf{u}}_n \\
\bar{M}_i &=& \frac{1}{\lambda} \int_{\Lambda}\int_{\Omega_i} \rho \;dV dt \\
\bar{M}_i \bar{\mathbf{u}}_i &=& \frac{1}{\lambda} \int_{\Lambda}\int_{\Omega_i} \rho \mathbf{u} \;dV dt 
\end{eqnarray}
Defining the mass fraction as $\bar{\omega}_i = \bar{M}_i / \bar{M} $ and
inserting these definitions into Eq. \ref{eq:energy-v}, we obtain
\begin{equation}
-\big[{ \bar{\omega}_w \bar{\mathbf{u}}_w + \bar{\omega}_n \bar{\mathbf{u}}_n  } \big]
\cdot  \nabla \big(\bar{\phi} \bar{p}\big)
= \frac{1}{\lambda V}  \int_{\Lambda}\int_{\Omega}
\bm{\tau} : \nabla \mathbf{u}\;dV dt \;.
   \label{eq:energy-vi}
\end{equation}
Since the right-hand side is positive, 
\begin{equation}
-\big[{ \bar{\omega}_w \bar{\mathbf{u}}_w + \bar{\omega}_n \bar{\mathbf{u}}_n  } \big]
\cdot  \nabla \big(\bar{\phi} \bar{p}\big)
     \ge 0 \;.
     \label{eq:2phase-inequality}
\end{equation}
Note that because the two
fluxes ${\omega}_i \bar{\mathbf{u}}_i$ have the same driving force, they cannot be treated independently. 
Nevertheless, since the driving force causes a corresponding response for each flux, 
\begin{equation}
    {\omega}_i \bar{\mathbf{u}}_i  
 = - \frac{ \mathbf{\mathsf{L}}_i}{\mu_i} \cdot
\nabla \big( \bar{\phi} \bar{p} \big) \;,
\label{eq:darcy-iii}
\end{equation}
where $ \mathbf{\mathsf{L}}_i$ is a phenomenological coefficient tensor for fluid 
$i\in\{w,n\}$. The choice to normalize the coefficient based on the 
viscosity $\mu_i$ is strictly conventional, i.e. to be consistent with the
expectation that $ \mathbf{\mathsf{k}}_i \rightarrow \mathbf{\mathsf{K}}$ as ${\omega}_i \rightarrow 1$. Note that $ \mathbf{\mathsf{L}}_i$ is not necessarily symmetric and positive definite,
since the constraint due to Eq. \ref{eq:2phase-inequality} applies only to their sum,
\begin{equation}
\sum_{i=w,n} 
      \mathbf{\mathsf{L}}_i \big[ \nabla \big( \bar{\phi} \bar{p} \big)\big]^2\ge 0 \;.
     \label{eq:effperm-constraint}
\end{equation}
If the fluids are perfectly immiscible and incompressible
then the mass fraction will simply track the fluid volume fraction based on the saturation
$\bar{s}_i$ and porosity $\bar{\phi}$, 
\begin{equation}
    \bar{s}_i \approx \omega_i \;, \quad
   \bar{\mathbf{q}}_i \approx  \bar{\phi} \bar{s}_i \bar{\mathbf{u}}_i
\end{equation}
where $\bar{\mathbf{q}}_i$ is the Darcy flux for fluid $i$. 
This expression is written as an approximation to account for compressible fluids since the mass flux does not necessarily coincide with the volumetric rate. The conventional relative permeability is defined such that
\begin{equation}
   \mathbf{\mathsf{L}}_i  =  k^r_i \mathbf{\mathsf{K}} \;,
\end{equation}
which will be satisfied in an isotropic material with constant porosity,
in which case Eq. \ref{eq:Darcy} can be recovered.

To obtain the conventional relative permeability relationship, the driving force must be the total pressure gradient and not the pressure gradient within a particular fluid. 
Many previous theoretical efforts have developed cross-coupling forms to describe
two-fluid flow in porous media \cite{whitaker1986flowb}. These forms present a significant
practical challenge, since additional coefficients are needed to close the resulting
system of equations. An advantage of Eq. \ref{eq:Darcy}  is that there fewer unknown coefficients are required and standard
experimental approaches can be applied to make inferences about the transport behavior.
We show that cross-coupling coefficients are unnecessary if the driving force is chosen
as described above, contingent on the average pressure taking the definition given in 
Eq. \ref{eq:pressure-definition}. We note that a cross-coupling form will be obtained if the pressure gradients 
are expanded as sums over each fluid using Eq. \ref{eq:subset}
\begin{equation}
    \nabla \big(\bar{\phi} \bar p\big)
     = \nabla \big(\bar{\phi} \bar{s}_w \bar{p}_w\big)+ \nabla \big(\bar{\phi} \bar{s}_n \bar{p}_n\big) \;.
     \label{eq:force-X-couple}
\end{equation}
At a basic level cross-coupling forms are a consequence of a choice to separately represent the driving
forces based on the pressures for each fluid. 
Defining the driving forces according to Eq. \ref{eq:force-X-couple} will lead to 
cross-coupling forms.  In this case, the total pressure gradient includes contributions from both fluids, each weighted by their respective volume fraction. 

\begin{figure*}
\centering
\includegraphics[width=1.0\linewidth]{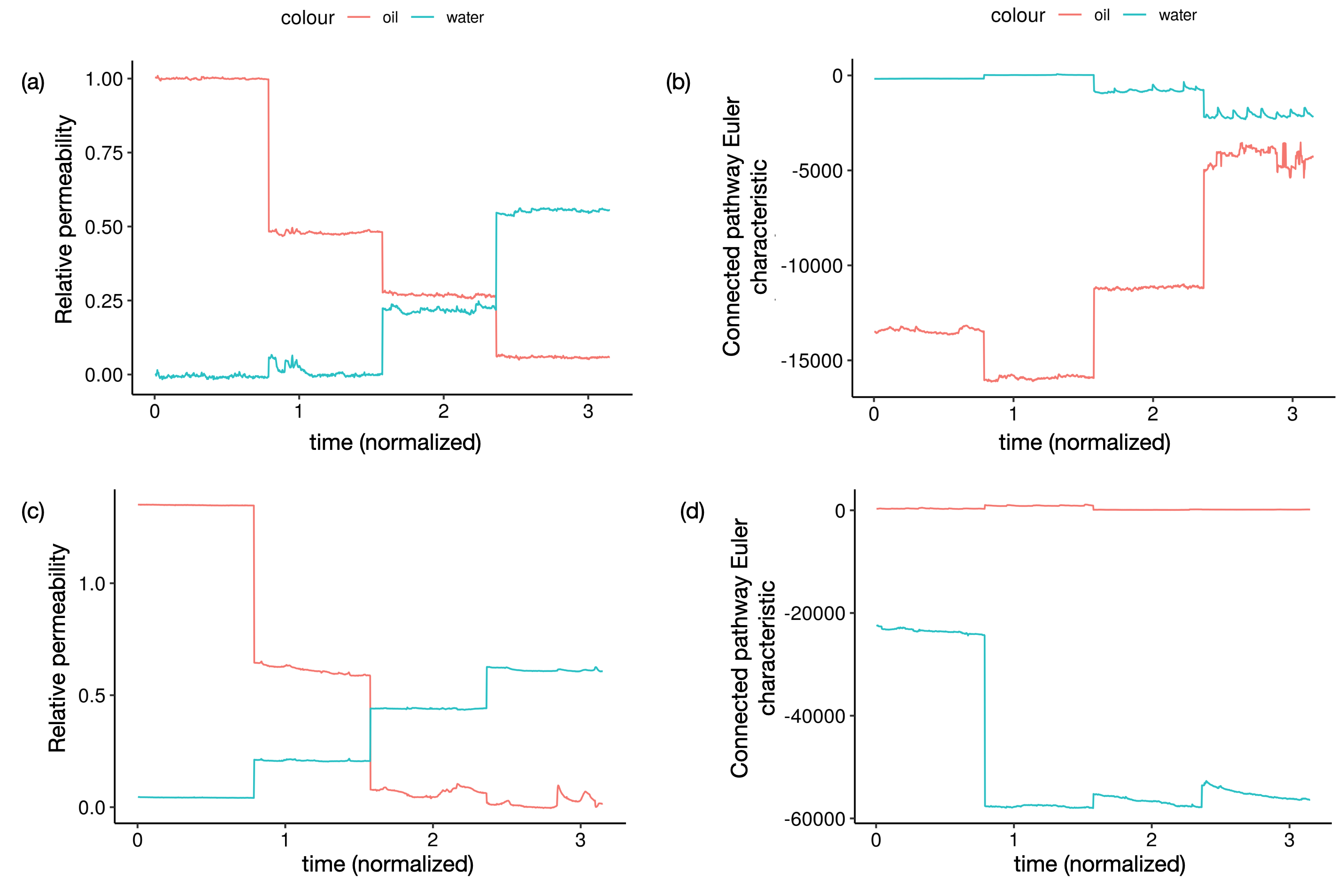}
\caption{
Simulation results showing the instantaneously measured relative permeability coefficients
as a function of simulation time based on water flooding in an oil-wet flow condition (drainage). 
(a) fluctuations are observed across all flow conditions due the transient response of the volumetric flow rate in response to pore-scale events; (b) the connected pathway Euler characteristic is demonstrated to vary significantly during flow at fixed saturation, indicating that structural changes to fluid connected pathways are relatively common.  }
\label{fig:relperm}
\end{figure*}

\section{Model Validation}

Steady-state simulations of immiscible fluid flow were performed using lattice Boltzmann method based on the color simulator in the LBPM software package \cite{mcclure2021lbpm}. A Bentheimer sandstone 
was used as the basis for simulations, collected at $1.66$ $\mu$m resolution \cite{Dalton_2019}. 
Simulations were initialized from a $1.494\mbox{ mm}\times1.494\mbox{ mm}\times2.656\mbox{ mm}$ sub-region of the original image with porosity $0.235$ and permeability $1850$ mD, for which steady-state fluid configurations had been generated previously \cite{Li_2020}. This was done to ensure that the maximum number of simulation timesteps were performed at steady state. Simulations were performed for both oil-wet and water-wet conditions to verify that the derived fluctuation criterion were met under a variety of flow conditions. Each simulation required approximately 10 hours of walltime using $72$ NVIDIA V100 GPU, and were performed using the Summit supercomputer. The capillary number for the displacement was $1\times 10^{-5}$. 


Instantaneously measured values for the relative permeability are reported in Figure \ref{fig:relperm}a. The instantaneous relative permeability can be considered as a time average where $\Lambda$ takes the value of a single simulation timestep, and the measured values fluctuate due to non-ergodicity. 
In this case, non-ergodicity is a consequence of the fact that the timescale required
for thermal information to propagate across $\Omega$ (i.e. thermal mixing) is longer than the timescale for $\Lambda$. The fluctuations can be removed if the energy dynamics are homogenized over a longer timescale. The flow behavior associated with Figure
\ref{fig:relperm} corresponds to a water-flooding process in a moderately oil-wet system, which can be considered as a drainage process, since water is the non-wetting fluid. Due to the contribution of Haines jumps during drainage, fluctuations to the fluid pressure signal are somewhat larger as compared to an imbibition process. Steady-state simulations are performed for four different saturation values, each showing the contribution of fluctuations with a clear plateau to indicate that stationary dynamics should be obtained if a sufficiently long timescale is considered. 

\begin{figure*}
\centering
\includegraphics[width=1.0\linewidth]{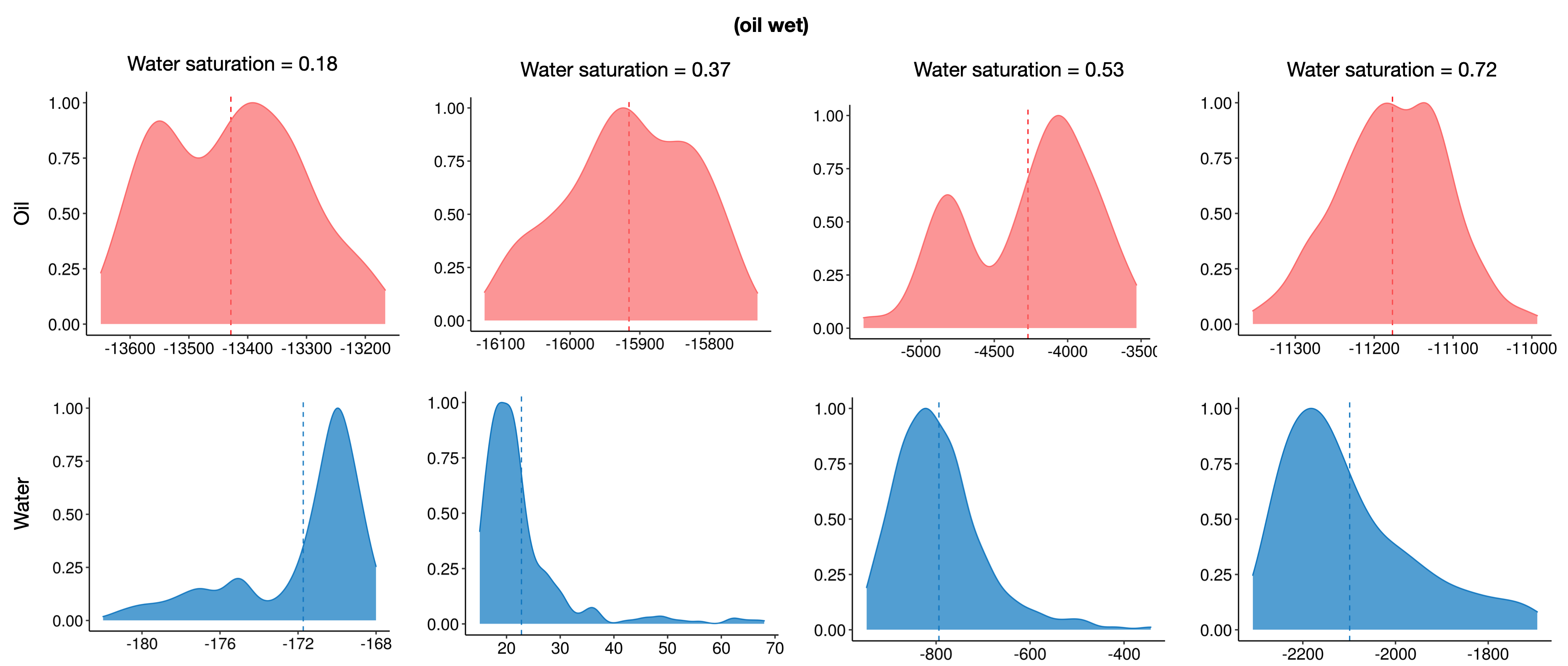}
\includegraphics[width=1.0\linewidth]{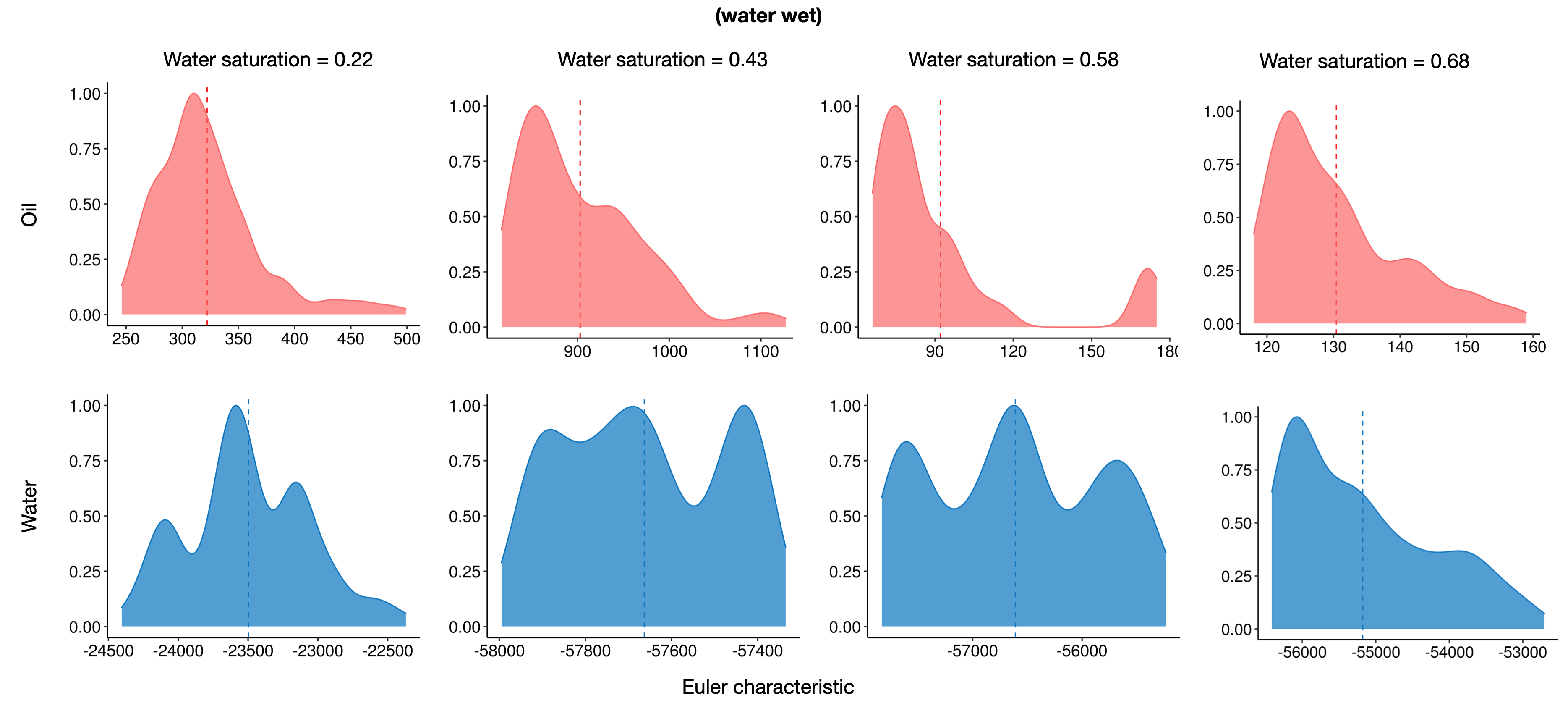}
\caption{Euler characteristic for the connected pathway extracted from the water and 
oil regions based on simulation cases shown in Figure \ref{fig:relperm}. The density
corresponds to the residence time associated with a particular connectivity state in the system. The dashed
line corresponds to the time-average value of the Euler characteristic. Transitions between these states will cause corresponding fluctuations to the energy modes. 
} 
\label{fig:intermittent}
\end{figure*}

The effects of intermittent connectivity can be assessed based on the Euler characteristic, which can be used to ascertain the topology of fluids within the system. The Euler characteristic for each fluid region $i$ is 
\begin{equation}
    \chi_{i} = \mathbb{N}_{i} - \mathbb{L}_{i} + \mathbb{C}_{i}\;,
\end{equation}
where $\mathbb{N}_{i}$ is the number of connected components,
$\mathbb{L}_{i}$ is the number of loops, and $\mathbb{C}_{i}$ is the number of cavities.
As long as there are no free solid particles such that a single connected solid structure
is formed along the grain contacts, cavities within fluid $n$ will be due to the connected components of fluid $w$, and vice versa. Figure \ref{fig:relperm}b shows that there are significant changes to connected pathway Euler characteristic for both fluids, which are most notable as the volume fraction for water increases and displaces oil from the primary flow pathways within the system. For each 
flow condition the fluid saturation is constant due to the periodic boundary condition used in the
simulation protocol (see \cite{mcclure2021lbpm} for a detailed description of the numerical implementation). This evidence suggests that intermittent connectivity is common during typical flow experiments, even those performed at low capillary numbers. 

\begin{figure*}
\centering
\includegraphics[width=1.0\linewidth]{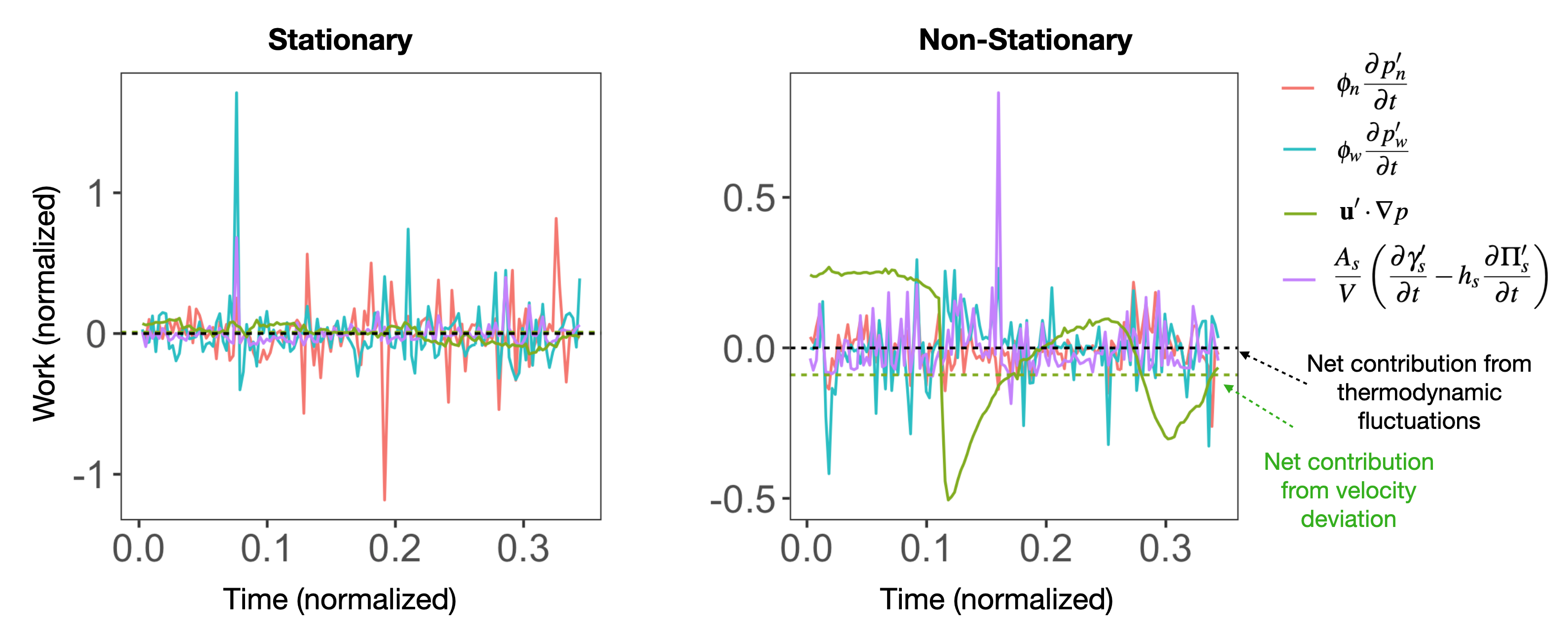}
\caption{
Contribution of all energy fluctuation terms for stationary and non-stationary displacements. 
For the stationary case (left), the net contribution due to work based on the sum of all fluctuations is 
effectively zero. For the non-stationary case (right), the net work due to velocity deviation 
is significant compared with the rate of dissipation in the system.
}
\label{fig:fluctuations}
\end{figure*}

Previous studies have demonstrated that fluid flow through permeable media can occur at intermittent flow pathways, where fluids can be periodically disconnected and reconnected through the snap-off and reconnection events \cite{reynolds2017pnas,spurin2019intermittent}. The continuous rearrangement of fluid interfaces at steady-state leads to a fluctuating rate of energy dissipation.  In this study, we demonstrate that the fluctuations based on the internal energy dynamics of the system in a stationary process will be averaged out and the net energy contribution to the system cancels due to the conservation of energy. The range of connectivity states explored within
each of the eight simulation cases  is presented in Figure \ref{fig:intermittent}. The density 
corresponds to the fraction of time that the system spends in each connectivity state. In 
many cases, multiple peaks are observed, suggesting that the fluid
connectivity is frequently toggling between several different states and that particular 
fluid structures are often unstable with respect to small disruptions in the flow behavior. 
By constructing the relative permeability based on averaging in both time and space, these various contributions can be incorporated into a single measurement based on the residence time that the system spends in each particular connectivity state.

The role of capillary events and intermittent connectivity can be further considered based on their impact on non-equilibrium fluctuations. 
Based on the arguments presented previously, stationary conditions should ensure satisfaction of the  constraints given by Eqs.~\ref{eq:stationary-constraint-i} and \ref{eq:stationary-constraint-ii}. Because the simulation domain is fully periodic, there is no net change in the energy due to boundary fluxes, and Eq.~\ref{eq:stationary-constraint-i}
will be satisfied. Since the lattice Boltzmann method is isothermal, the energy fluctuation due to temperature heterogeneity is identically zero,
\begin{equation}
    \Big< S \frac{\partial T^\prime}{\partial t} \Big> = 0\;.
\end{equation}
Subject to this restriction, for immiscible fluid flow through porous media, the stationary constraint given in Eq.~\ref{eq:fluctuation} can be expressed in the alternative form (see McClure et al. for complete details \cite{mcclure2020capillary})
\begin{equation}
   \Big<\mathbf{u}^\prime\cdot \nabla p + {\frac{A_s}{V} \left( \frac{\partial \gamma_s^\prime }{\partial t} - h_s \frac{\partial \Pi_s^\prime }{\partial t} \right)   } - \phi_w \frac{\partial p_w^\prime }{\partial t} - \phi_n \frac{\partial p_n^\prime }{\partial t} \Big> = 0
   \label{eq:2phase-darcy-fluctuation}
\end{equation}
where $p_i$ is fluid pressure for $i \in \{w, n\} $, $\gamma_s$ is fluid-solid surface energy along the solid material, and $\Pi_s $ is disjoining pressure, $\phi_i$ is fluid volume fraction for $i \in \{w, n\} $, $A_s$ is surface area for the solid material, and $h_s$ is the film thickness along the solid material. This form is obtained when the internal energy is presumed to depend on an alternate set of extensive
variables based on position of the Gibbs dividing surface in the system. The two representations are necessarily equivalent. 
Within the simulation, the wetting energy and fluid pressures are computed directly and are are numerically integrated to obtain averages. Time derivatives for time-and-space averaged quantities are determined as described in Appendix B.

\begin{figure*}
\centering
\includegraphics[width=1.0\linewidth]{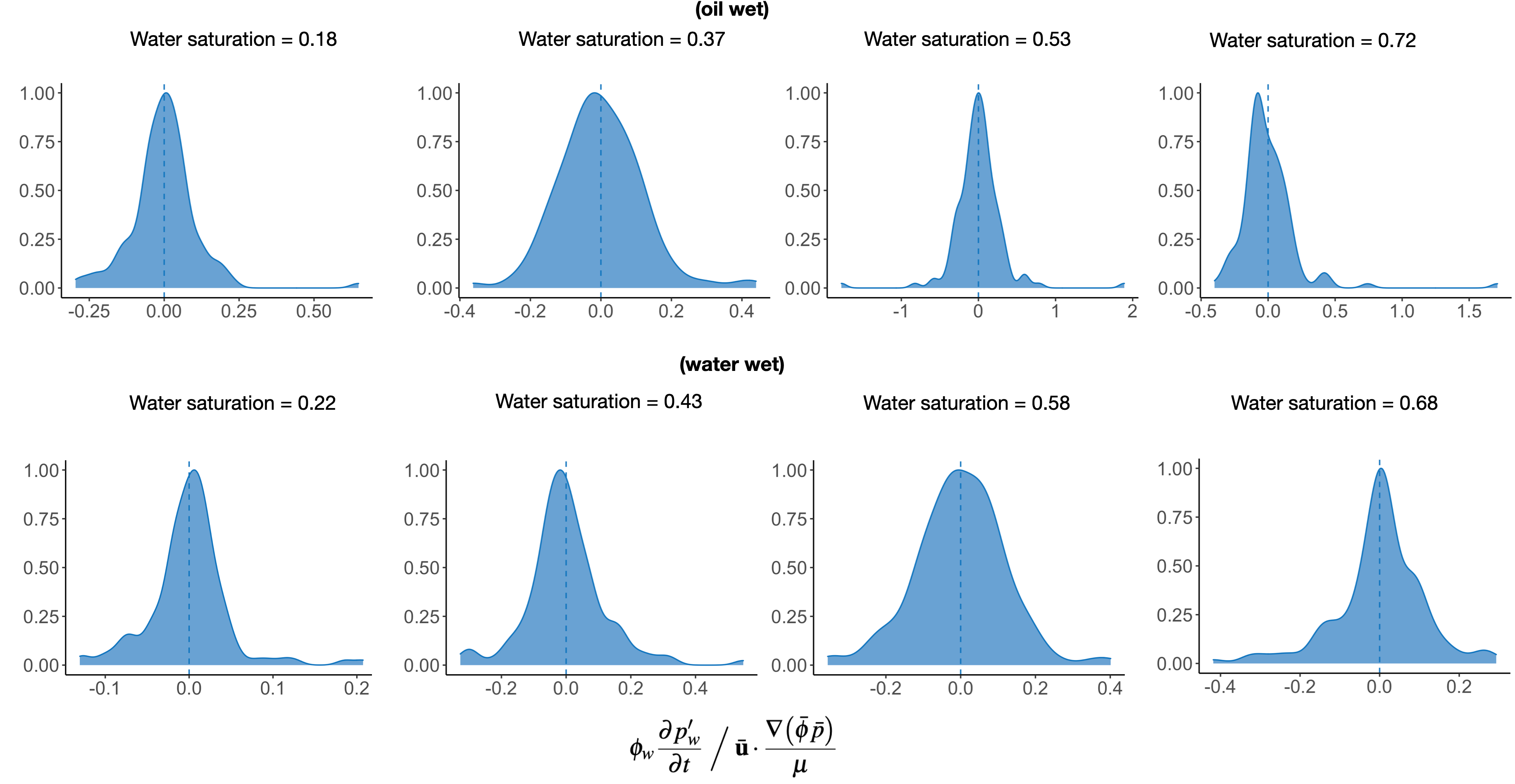}
\caption{
Normalized work due to fluctuations associated with the water pressure. 
Net contributions are expected to be more significant in situations where both the pressure and the water volume fraction are changing with time.
}
\label{fig:fluctuations-pw}
\end{figure*}

\begin{figure*}
\centering
\includegraphics[width=1.0\linewidth]{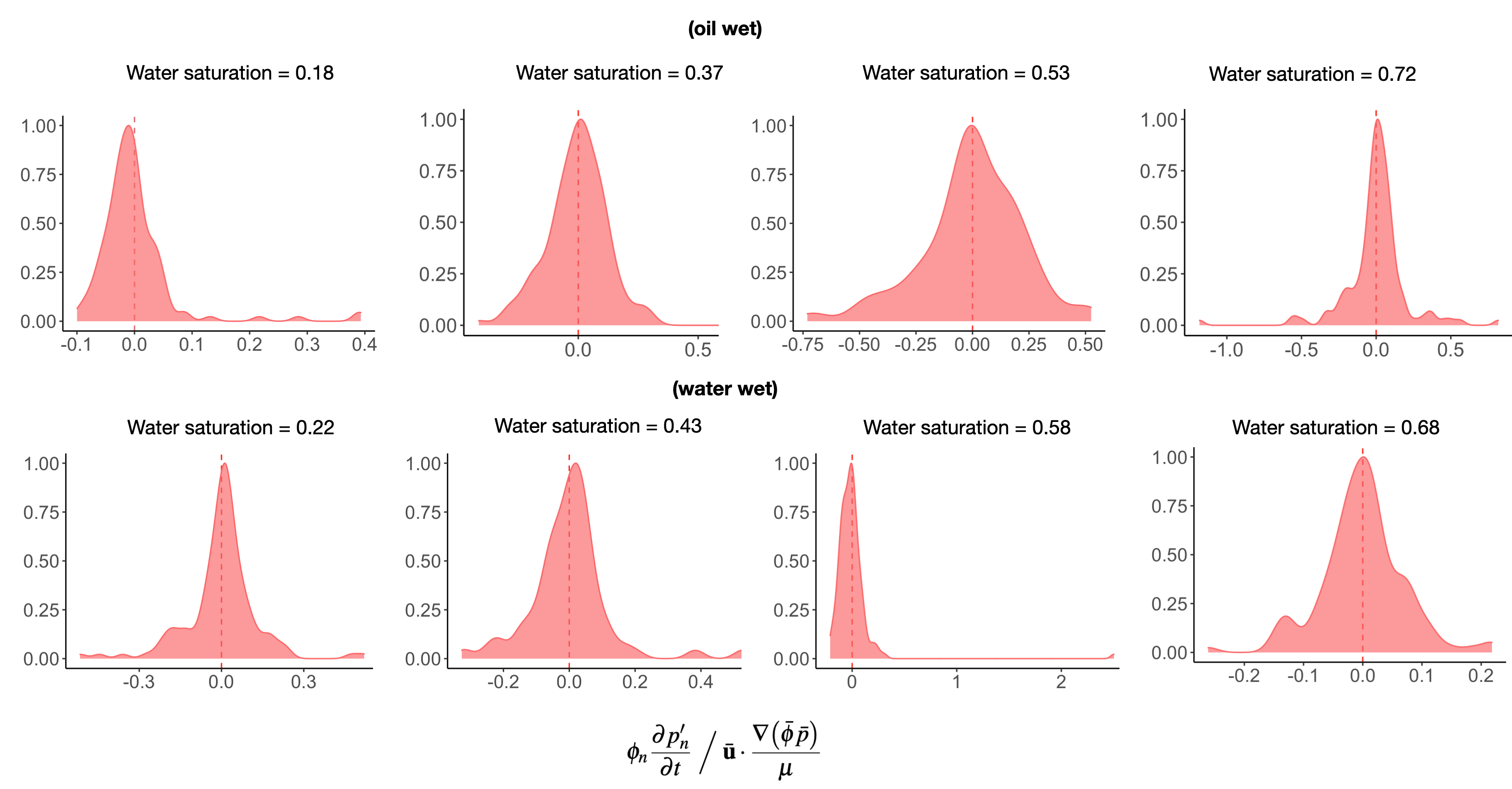}
\caption{
Normalized work due to fluctuations associated with the oil pressure. 
Net contributions are expected to be more significant in situations where both the pressure and the oil volume fraction are changing with time.
}
\label{fig:fluctuations-pn}
\end{figure*}

Defining relative permeability as a stationary process leads to the fluctuation constraint stated in Eq. \ref{eq:2phase-darcy-fluctuation}.  If this expression does not hold, then the flow process cannot be considered to be stationary. This distinction is illustrated in Figure \ref{fig:fluctuations-pn},
which shows each contribution to energy fluctuations for $\lambda=100,000$ simulation timesteps. The energy scale is normalized based on the dissipated heat, which is given by Eq. \ref{eq:darcy-i}.
As with typical multiphase flow experiments, the energy associated with the system dynamics is comparable to the overall rate of dissipation in the system. However, the net contribution of 
of these effects can be effectively removed in a stationary system, due to cancellation of the contributions over a sufficiently long timescale. A non-stationary result is obtained when 
particular aspects of the dynamics occur at a slow timescale, such that net work is performed by
a particular mechanism is non-zero over the time averaging window $\lambda$. In this context the integral in Eq. \ref{eq:2phase-darcy-fluctuation} provides a quantitative assessment of the extent to which a particular process can be considered to be stationary. If the net work done by fluctuation terms
is significant compared to the rate of dissipation in the system, the underlying process should not be considered to be stationary. A relative permeability coefficient determined based on a non-stationary process may be inaccurate due to the fact that it is uncertain whether energy stored in the fluctuation terms will contribute to mass transfer or if this energy will ultimately be dissipated. 

Based on simulation data, we can accurately characterize the essential contributions to the 
fluctuating energy dynamics. Figures \ref{fig:fluctuations-pw} and \ref{fig:fluctuations-pn} show the distribution of energy fluctuations for the water and oil fluid pressures. 
 We can observe that pressure fluctuations may be non-Gaussian, multi-modal and asymmetric. At infinitely large timescales, the fluctuations should approach a Gaussian distribution. However, this is not necessarily the case for typical experimental timescales, and significant fluctuations are observed at the timescale to take pressure measurements \cite{mcphee2015core}. For the timescale considered here, non-Gaussian distributions are most likely caused by the pore-scale displacement events associated with transitions between different connectivity states. These fluctuations are the expressions of internal energy exchanges, where capillary energy is converted into kinetic energy, with frequent jumps and topology changes due to the coalescence and snap-off events between fluid clusters. Due to the significant difference in energy fluctuation distributions across the whole system, the changes of energy fluctuation distributions depend on the wetting condition and fluid saturation. The structure of fluctuations is very sensitive to the saturation condition, since the frequency of pore-scale events depends on the pore occupancy. Even though the distributions are not Gaussian, the net contribution to the system energy dynamics is
 approximately $4$--$6$ orders of magnitude smaller than the overall rate of energy 
 dissipation. For the simulations considered here, only very small variations occur for the fluid volume fractions. This is significant because volume is the extensive variable 
 linked with pressure in the thermodynamic representation. As shown in Appendix A, 
 the net contribution for an energy fluctuation will be close to zero if the extensive variable is nearly constant with time. However, the more the extensive variable is  fluctuating, the greater the fluctuation can contribute to the energy dynamics. This is significant because fluctuations to the fluid volume fraction are much more significant
 in an experimental setting \cite{rucker2021}. Experimental evaluation of these terms should therefore be treated as an important priority for further validation of the theory.

\begin{figure*}
\centering
\includegraphics[width=1.0\linewidth]{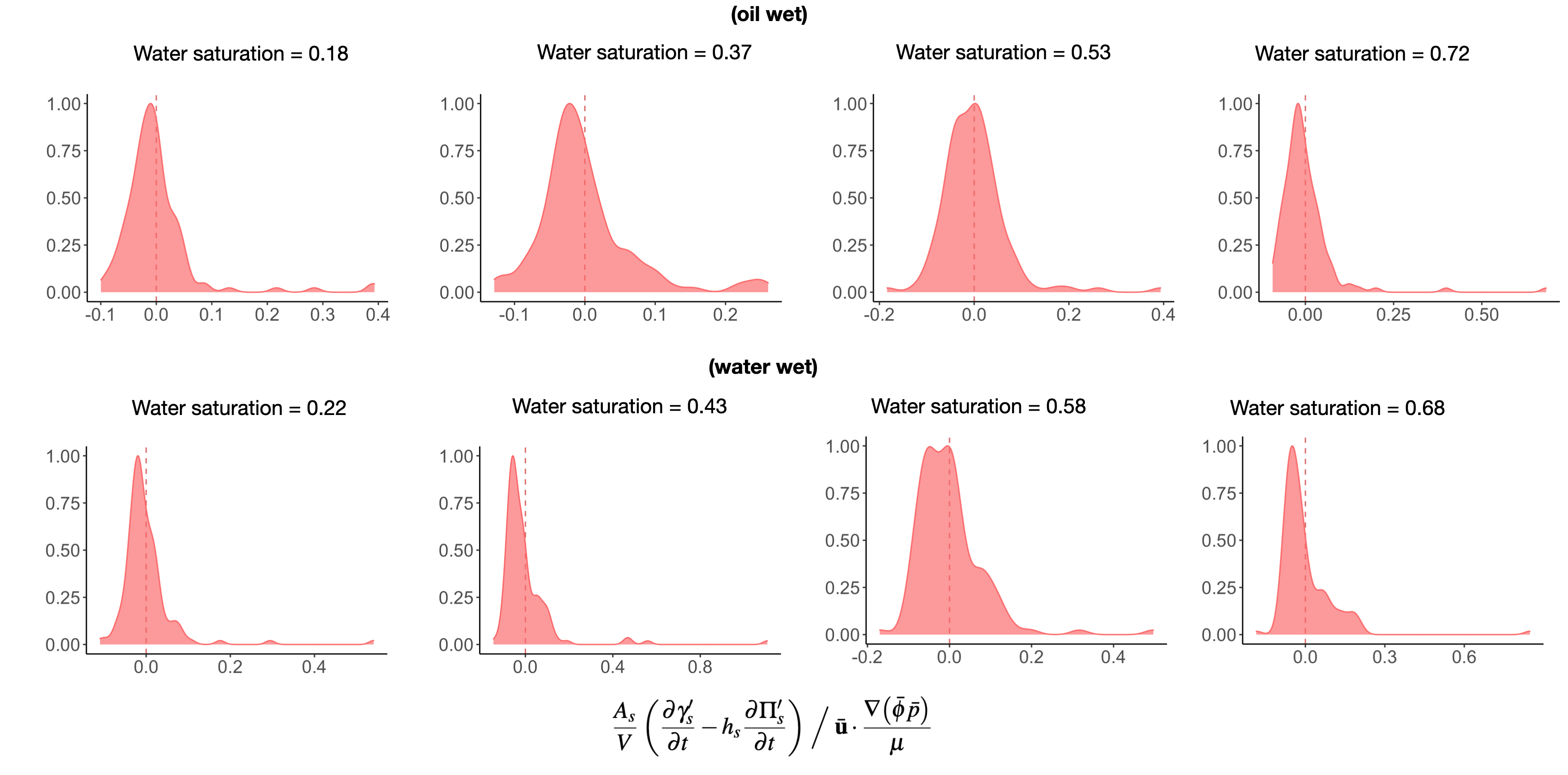}
\caption{ Normalized work due to fluctuations associated with solid surface wetting energy. Net contributions will become important when the solid surface area is
changing with time.
}
\label{fig:fluctuations-wetting}
\end{figure*}

Figure \ref{fig:fluctuations-wetting} shows the normalized work associated with fluctuations
in the surface wetting energy. The mechanisms that cause these fluctuations are due to
film swelling and movement of the contact line. The total surface wetting 
energy is constructed as an integral that includes surface excess energy due to interactions
that occur over the fluid-solid surface, which has area $A_s$
\cite{mcclure2021lbpm}. Again referring to the calculation in Appendix A, this contribution will
be zero if the solid surface area is constant. Note that this does not mean that energy does not
accumulate at the surface due to the redistribution of fluids; the average surface
tension $\bar{\gamma}_s$ will be altered in such a scenario, such that the associated energy is 
accounted for based on $\bar{\gamma}_s A_s$. Surface wetting fluctuation terms will contribute 
when $A_s$ is not constant over $\Lambda$.

\begin{figure*}
\centering
\includegraphics[width=1.0\linewidth]{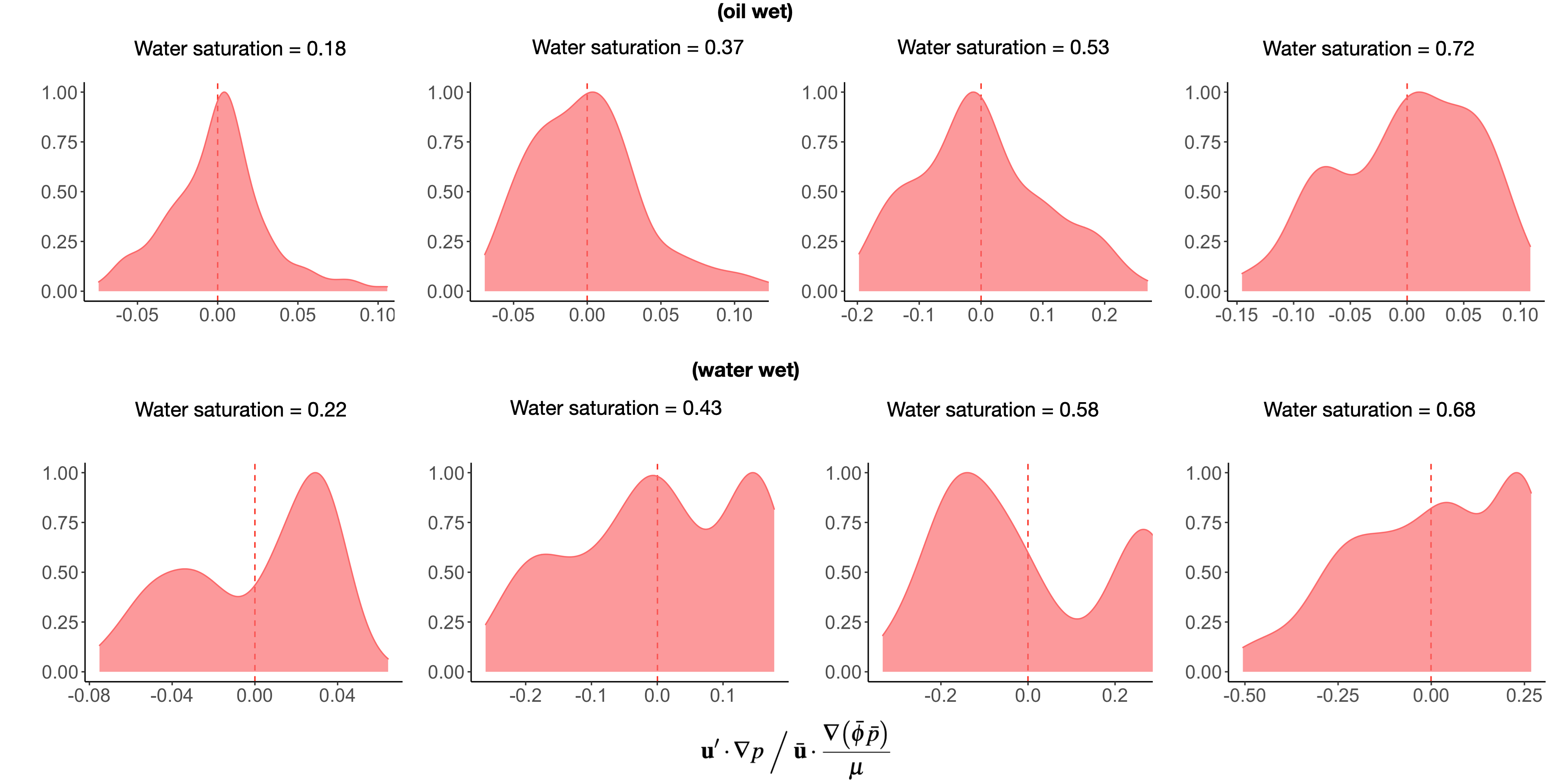}
\caption{ Normalized work due to mechanical fluctuations in the velocity and pressure.
Net contributions are expected to become important when there are more frequent pore-scale displacement events. 
}
\label{fig:fluctuations-velocity}
\end{figure*}

For the simulations considered here, mechanical fluctuations due to $\mathbf{u}^\prime\cdot \nabla p$
are the most significant contribution to the energy dynamics, which are due to pore-scale events.
The distribution for these contributions are depicted in Figure \ref{fig:fluctuations-velocity}.
This term may be understood based on the work of 
Berg et al. \cite{PhysRevE.102.033113}. Considering the situation
where the fluids are incompressible, the volume integral can be 
treated as a surface integral over the fluid-fluid meniscus $\Gamma_{wn}$
(see Appendix B)
\begin{eqnarray}
\int_{\Lambda} \int_{\Omega}
 \nabla \cdot \big(p \mathbf{u}^\prime\big) \; dV dt
 &=& 
\int_{\Lambda}  \int_{\Gamma_{wn}}
  \big(p_w - p_n\big) \mathbf{u}^\prime \cdot \mathbf{dA} dt\;.
\label{eq:B-ii}
\end{eqnarray}
In other words, the term accounts for the energy dynamics due to mechanical work performed during rearrangements of the fluid-fluid interfaces. A pore-scale event such as a Haines jump will cause both the local fluid pressure and the flow velocity to fluctuate. If the fluid volume fractions are constant, the timescale for these 
surface contributions will tend to be the dominant factor for obtaining stationary 
conditions. Since the fluid volume fraction changes based on interface movement,
this term will also contribute significantly in any situation
where fluid volume fractions change. 

\section{Conclusions}

In this work we have shown that the standard expression for relative permeability can be derived
from conservation of energy using averaging in both time and space and assuming that the dynamics are stationary. In conventional homogenization
theories, it is generally assumed that the contribution of capillary fluctuations will be negligible at large length scales based on the ergodic hypothesis. 
Our approach explicitly considers the timescale needed to obtain stationary conditions for 
a particular system, meaning that relative permeability coefficients can be obtained that accurately represent the rate of energy dissipation.
We mathematically define a stationary process as one for which there can be no net
change to extensive measures of the system state when considered over the timescale of the average.
Furthermore, there can be no accumulation of energy based on the rate of work and heat exchange
at the boundary of the system. Due energy conservation, energy fluctuations defined based on classical thermodynamics must also cancel, which is required to observe detailed balance in the macroscopic system. Subject to these constraints, a flux-force form is obtained that can be used to derive the
conventional relative permeability relationship. This derivation shows that the thermodynamic 
driving force should be the total pressure (potential) gradient, which is a simple average 
over the considered region of time and space. We further show that if the spatial region is 
segregated into sub-regions associated with the wetting and non-wetting fluids, cross-coupling forms
become necessary. Previous derivations that have arrived at cross-coupling forms as the general
approximate form for momentum transport are a consequence of this spatial subdivision. Compared with
cross-coupling forms, fewer unknowns are required by the conventional relative permeability, representing a significant practical advantage in terms of measurement. We conclude that there is no
disadvantage to using the conventional relative permeability; provided that the driving force
is appropriately identified the relative permeability coefficients will accurately predict the rate
of energy dissipation in the system. 

Results derived in this work do not assume that the flow behavior is ergodic at large length scales. 
Instead, ergodicity is assumed at small length scales, and a fluctuation criterion is established
to evaluate the proper scale of a time average. For a stationary process, the contribution of fluctuations can be entirely removed provided that a sufficiently long averaging timescale is used. Using simulation data performed at low capillary number ($\mathsf{Ca}\sim1\times10^5$) for multiple different saturation and 
wetting condition, we show that individual fluctuations are not necessarily symmetric or 
Gaussian. However, it is sufficient for the sum of the fluctuations to cancel, since this implies
that the associated thermodynamic modes do not perform net work on the system over the timescale
for the average. This timescale is physically linked to the effect of dynamic connectivity, which is assessed based on the Euler characteristic for the connected pathway for each fluid. Our analysis suggests that 
typical multiphase flows permute through a variety of connectivity states, each of which
is likely to be unstable. The time average can be thought of as accounting for residence time that
the system spends in each particular connectivity state. Since the Euler characteristic is an observable extensive measure of the system state, we assert that this topological measure should 
be included when making a determination as to  whether or not flow behavior meets the stationary requirements derived in this work. An important area for future work are
experimental studies that consider how fluctuations to the fluid volume fraction 
impact the the internal energy dynamics. 

\section*{Acknowledgements}
This research used resources of the Oak Ridge Leadership Computing Facility at the Oak Ridge National Laboratory, which is supported by the Office of Science of the U.S. Department of Energy under Contract No. DE-AC05-00OR22725.
R. T. A. acknowledges Australian Research Council Future Fellowship (FT210100165). The authors would like to thank Signe Kjellstrup, Dick Bedeaux and Dag Chun Stadnes for useful discussions. 

\appendix 
\section{Evaluation of Energy Fluctuation Terms}

To evaluate the fluctuation terms in Eq. \ref{eq:2phase-darcy-fluctuation},
the derivative of the macroscopic pressure is needed. For averages of intensive measures $Y_i$ with associated extensive variable $X_i$, the derivative is determined from
\begin{eqnarray}
    \frac{\partial{\bar{Y}_i}}{\partial t}
    &=& \frac{1}{\bar{X}_i} \Bigg[\frac{\partial}{\partial t} \Big( \bar{Y}_i\bar{X}_i\Big)
    - \bar{Y}_i \frac{\partial \bar{X}_i }{\partial t} \Bigg]\;.
    \nonumber \\
     &=& \frac{1}{\bar{X}_i} \Bigg[\frac{\partial}{\partial t} \Big< {Y}_i{X}_i\Big>
    - \bar{Y}_i \frac{\partial \big<{X}_i\big> }{\partial t} \Bigg]\;.
    \label{eq:derivative}
\end{eqnarray}
The quantities in brackets can be easily obtained, since the derivative of a definite integral is given by the fundamental theorem of calculus
\begin{equation}
    \frac{\partial}{\partial t} \int_{t_0}^{t_1} A(\tau) d\tau 
    = A ({t_1})- A (t_0)\;.
\end{equation}
Note that the fluctuation term for an intensive property $Y_i$ will be zero if the associated extensive conjugate property $X_i$ is constant with time.
The energy fluctuation is expanded as
\begin{eqnarray}
    \Big<  X_i \frac{\partial Y_i^\prime}{\partial t} \Big>
    = \Big<  X_i 
    \Big(\frac{\partial Y_i}{\partial t}  - 
    \frac{\partial \bar{Y}_i}{\partial t} \Big) \Big>\;.
    \label{eq:A-fluctuation}
\end{eqnarray}
Now suppose that ${X}_i = \bar{X}_i$, which do not depend on time 
\begin{equation}
    \frac{\partial}{\partial t} \Big <X_i Y_i \Big> = \frac{\partial (\bar{X}_i \bar{Y}_i)}{\partial t}  = 
     \Big< {X}_i \frac{\partial \bar{Y}_i }{\partial t}\Big>  = 
    \Big<{X}_i \frac{\partial Y_i}{\partial t}\Big>\;.
\end{equation}
In this situation Eq. \ref{eq:A-fluctuation} is zero. The energy
fluctuation terms thereby only contribute when both the extensive measure
and the associated intensive conjugate measure are co-fluctuating. 

\section{Evaluation of Velocity Deviation}

Previous work of Berg et al. has considered geometric contributions
to the rate of work \cite{PhysRevE.102.033113}. We consider a similar approach, but focused instead on fluctuations due to the velocity deviation.
In the event that the flow is incompressible, we may use the fact that
$\nabla \cdot \mathbf{u} = 0$ to write
\begin{eqnarray}
\int_{\Omega}
\mathbf{u}^\prime\cdot \nabla p \; dV  = 
\int_{\Omega}
 \nabla \cdot \big(p \mathbf{u}^\prime\big) \; dV \;. 
\label{eq:B-i}
\end{eqnarray}
The volume integral can then be divided into the contributions over the 
wetting and non-wetting fluids, which are then transformed to surface integrals based on the Gauss divergence theorem,
\begin{eqnarray}
\int_{\Omega}
 \nabla \cdot \big(p \mathbf{u}^\prime\big) \; dV 
 &=& \int_{\Omega_n}
 \nabla \cdot \big(p \mathbf{u}^\prime\big) \; dV +
 \int_{\Omega_w}
 \nabla \cdot \big(p \mathbf{u}^\prime\big) \; dV\;,
 \nonumber \\
&=& 
 \int_{\Gamma_{wn}}
  \big(p_w - p_n\big) \mathbf{u}^\prime \cdot \mathbf{dA}\;,
\label{eq:B-ii}
\end{eqnarray}
where $\Gamma_{wn}$ is the meniscus between the two fluids. 
Note that because the simulations used to develop our results are fully periodic, there is no external boundary in the system.  
To carry out the time integration numerically, 
we form the following approximation for the integrand
\begin{eqnarray}
 \int_{\Gamma_{wn}}
  \big(p_w - p_n\big) \mathbf{u}^\prime \cdot \mathbf{dA}\;
  \approx (\tilde{p}_w - \tilde{p}_n) 
  (\tilde{\mathbf{u}} - \bar{\mathbf{u}}) \tilde{A}_{wn} \;,
\label{eq:B-iii}
\end{eqnarray}
where the instantaneous spatial averages are 
\begin{eqnarray}
 \tilde{p}_i  &=&  \int_{\Omega_i} p \; dV\;, \\
 \tilde{A}_{wn}  &=&  \int_{\Gamma_{wn} }  dA, \\
 \tilde{\rho} \tilde{\mathbf{u}} &=&  \int_{\Omega} \rho \mathbf{u} \; dV.
\label{eq:B-iv}
\end{eqnarray}
This approximation is based on several assumptions:
(1) deviations in the average flow velocity are distributed similarly
to the deviations in the average flow velocity in at the meniscus;
and (2) the pressure difference at the meniscus is well approximated by
the difference between the volume average pressures. While these assumptions are not expected to hold exactly, it is noted that direct experimental measurement for the left hand side of Eq. \ref{eq:B-iii} is non-trivial. 
The form suggested by the right-hand side is more straightforward to obtain,
and likely to provide a reasonable approximation in most situations. 
Time averages may then be obtained by integrating the spatial averages $\tilde{p}_i(t)$,$\tilde{\mathbf{u}}(t)$ and $\tilde{A}_{wn}(t)$.

\bibliography{References}

\end{document}